\documentclass{template}

\usepackage{amsmath}
\usepackage[version=3]{mhchem} % Formula subscripts using \ce{}
 \usepackage{setspace}
 \usepackage{siunitx}
\usepackage[T1]{fontenc}

\usepackage{helvet}

\begin{document}

\title{Sol-gel Barium Titanate Nanohole Array as a Nonlinear Metasurface and a Photonic Crystal}

\maketitle

\author{Ülle-Linda Talts*[1]},
\author{Helena C. Weigand[1]},
\author{Grégoire Saerens[1]},
\author{Peter Benedek[2]},
\author{Joel Winiger[3]},
\author{Vanessa Wood[2]},
\author{Jürg Leuthold[3]},
\author{Viola Vogler-Neuling[1,4]},
\author{Rachel Grange[1]}

\begin{affiliations}

\unit{[1]} Optical Nanomaterial Group, Institute for Quantum Electronics, Department of Physics,ETH Zurich, HPT H10, Auguste-Piccard-Hof 1, 8093 Zurich, Switzerland\\
Email Address: utalts@phys.ethz.ch\\ 
\unit{[2]} Institute for Electromagnetic Fields, Department of Information Technology and Electrical Engineering, ETH Zurich, ETZ H96, Gloriastrasse 35,8092 Zurich, Switzerland\\
\unit{[3]} Institute for Electronics, Department of Information Technology and Electrical Engineering, ETH Zurich, ETZ H96, Gloriastrasse 35,8092 Zurich, Switzerland\\
\unit{[4]} Soft Matter Physics Group, Adolphe Merkle Institute, University of Fribourg, Chemin des Verdiers 4, Fribourg, 1700, Switzerland\\
\end{affiliations}

\begin{onehalfspace}
\justifying

%\begin{abstract}

The quest of a nonlinear optical material that can be easily nanostructured over a large surface area is still ongoing. Here, we demonstrate a nanoimprinted nonlinear barium titanate 2D nanohole array that shows optical properties of a 2D photonic crystal and metasurface, depending on the direction of the optical axis. The challenge of nanostructuring the inert metal-oxide is resolved by direct soft nanoimprint lithography with sol-gel derived barium titanate enabling critical dimensions of 120 nm with aspect ratios of 5. The nanohole array exhibits a photonic bandgap in the infrared range when probed along the slab axis while lattice resonant states are observed in out-of-plane transmission configuration. 
The enhanced light-matter interaction from the resonant structure enables to increase the second-harmonic generation in the near-UV by a factor of 18 illustrating the potential in the flexible fabrication technique for barium titanate photonic devices.

%\end{abstract}

\section{Introduction}

Nonlinear optics has greatly expanded the field of photonics with phenomena such as frequency conversion processes and quantum light generation. While current applications of these photon-photon interactions rely primarily on high-intensity optical sources and bulky crystals, new strategies are being developed to enhance the intrinsically weak higher-order nonlinear effects.
Combining nanostructures with a nonlinear optical medium has demonstrated to increase the efficiency of parametric quantum processes such as second harmonic generation (SHG) \cite{Timpu2019}, sum-frequency generation \cite{Morales2022}, four-wave mixing \cite{Jin2016} or spontaneous parametric down-conversion \cite{SantiagoCruz2021}. 

Several material platforms support these nonlinear processes yet certain drawbacks have limited their applicability. III-V materials have strong nonlinear coefficients but signals can be screened by the substrate required for epitaxial growth or the second-order nonlinear optical susceptibility tensor is oriented unfavorably for out-of-plane excitation \cite{Adachi2009, Menon2022}. Poled polymers show an impressive modulation efficiency with an order of magnitude larger electro-optic coefficient compared to lithium niobate, but the material stability is hindering its use in long-term applications \cite{Ullah2021, Taghavi2022}. Plasmonic structures permit the highest electric field confinement but they are subject to high optical losses and low damage threshold \cite{Boyd2008}.

Non-centrosymmetric metal-oxides, firstly only available as bulk crystals, show rather large second-order susceptibility coefficients, high damage threshold and a wide transparency range. These properties are exploited for optical modulators and laser gain media \cite{Karvounis2020}. The recent advances in fabrication of lithium niobate on insulator devices have demonstrated the potential in combining these metal-oxide thin films with advanced nanofabrication for miniuatrized and efficient optical devices such as high-speed electro-optic modulators and on-chip optical parametric oscillators, to name a few \cite{Boes2023}. Also other metal-oxides such as barium titanate \cite{Karvounis2020}, barium borate \cite{Solntsev2021}, potassium niobate \cite{Dues2022} and potassium titanyl phosphate \cite{Kores2021} have the potential to be an alternative nonlinear optical material platform. However, the limited thin film availability and the transferability of CMOS nanofabrication processes make it challenging, motivating the search for novel nanofabrication approaches. 

We choose to focus on barium titanate (BaTiO\textsubscript{3}, or BTO), a metal-oxide with intrinsic nonlinear optical processes, characterized by a second-order susceptibility tensor of d\textsubscript{15}=15.7 \unit{pm\raiseto{-1}V} in thin film, energy gap of 3.5 \unit{eV} (low optical material losses above 354 \unit{nm}), ultra-high  electro-optic response of r\textsubscript{15}=1300 \unit{pm\raiseto{-1}V} in the unclamped case and physical resilience to mechanical as well as optical damage \cite{Karvounis2020}. Due to the chemical inertness, insulating nature and no commercial availability of high quality thin film wafers,nanofabrication of the metal-oxide photonic crystals (PhC) and metasurfaces has been challenging using a similar nanolithography approach as for semiconductor devices \cite{Weigand2021, Girouard2017}. 
Molecular beam epitaxy deposited BTO thin films have been used for active photonic integrated circuits. Due to the limited thickness of the achieved films and challenging etching, hybrid structures with other semiconductors are needed to form light guiding films \cite{Girouard2017, Eltes2020}.  
 
In a previous publication, we have reported BTO metasurfaces with 15 times increased SHG in the near-UV range by a hybrid device with a SiN\textsubscript{x} capping layer to form the resonant nanodisk arrays \cite{Timpu2019}. To achieve both previously mentioned structures, focused ion beam milling or electron-beam lithography with plasma etching was employed, requiring highly specialized and optimized tools. Here we report on an alternative fabrication technique to circumvent the need for hybrid structures as well as the difficulty to pattern and etch the chemically resistant and insulating metal-oxide.

Bottom-up fabrication via direct soft nanoimprint lithography (SNIL) has the precision of top-down fabrication and the flexibility to be applied to novel materials without developing a new etching recipe \cite{Modaresialam2021}. Most commonly this technique uses polydimethylsiloxane (PDMS) stamps, fabricated as an inverse from a nanostuctured Si-based master mold. This enables imprinting the desired structures into the material of interest. It is an established method for patterning photoresists for top-down fabrication, but directly imprinting nanoparticles and sol-gel solutions has been shown as another route for highly-scalable nanopatterning of metal-oxides \cite{Modaresialam2021}.

Nanoparticles of BTO above 30 \unit{nm} have similar crystal properties as the bulk crystal \cite{Xiao2008} and since they can be dispersed in a solvent, molding the particles into periodic structures is possible \cite{Vogler-Neuling2020}. However, due to the limited packing density and inherent rough surfaces of nanoparticle assemblies, achieving resonant photonic structures has been difficult due to scattering. An alternative to nanoparticle solutions are sol-gels, which form a polymer-like metal-oxide network in a solution and becomes a dense xerogel thin film after annealing \cite{Edmondson2020}. Lithium niobate nanohole arrays have been made by capillary imprinting an aqueous sol-gel and the authors observed SHG enhancement when combined with gold particles \cite{Alarslan2022}. Effective electric field confinement in nanoimprinted polycrystalline PhC has shown to enable quasi-CW lasing in metal halide perovskite, but similar increased light-matter interaction has not been shown in bottom-up fabricated nonlinear metal-oxide structures \cite{Moon2022}.  

Sol-gel derived BTO 2D PhC formed a photonic bandgap\cite{Hirano2005} and while the investigation of this metal-oxide's electro-optic properties established a competitive value with thin film lithium niobate \cite{Edmondson2020}, the optical nonlinear properties are still largely unexplored. 

Because high quality PhC underly devices such as high Purcell factor cavity resonators \cite{Altug2006, Li2019, Diziain2013}, dispersionless and low loss waveguides \cite{Krauss2007} or topological insulators \cite{Tang2022}, we choose a design of a hexagonal nanohole array in solution-processed BTO slab. Air holes resonant at specific wavelengths have also been realized for low-loss Mie-resonant cavities \cite{Hentschel2023} and sensors with high energy confinement at metasurface-air interface \cite{Conteduca2021}. Similarly, nanohole arrays in a nonlinear optical material slab, such as BTO, increase the light-matter interaction and SHG \cite{Timpu2019,Li2019,Diziain2013}. Using direct SNIL to pattern the sol-gel BTO, we achieve high aspect ratios of 5 and minimized scattering from the smooth side-walls. The in-plane and out-of-plane light interaction in the bottom-up fabricated structures is characterized by measuring a slab photonic crystal bandgap and Mie surface lattice resonances, respectively. Additionally, we show that the sol-gel derived films have nonlinear optical properties which can be enhanced by the nanostructures indicating this technique can be used to increase the nonlinear light generation or electro-optic modulation efficiency.

\section{Results and Discussion}
\subsection{Barium Titanate Sol-gel and Nanoimprinting}

\begin{figure}[h]
%\centering
\centering\includegraphics[width=1\textwidth]{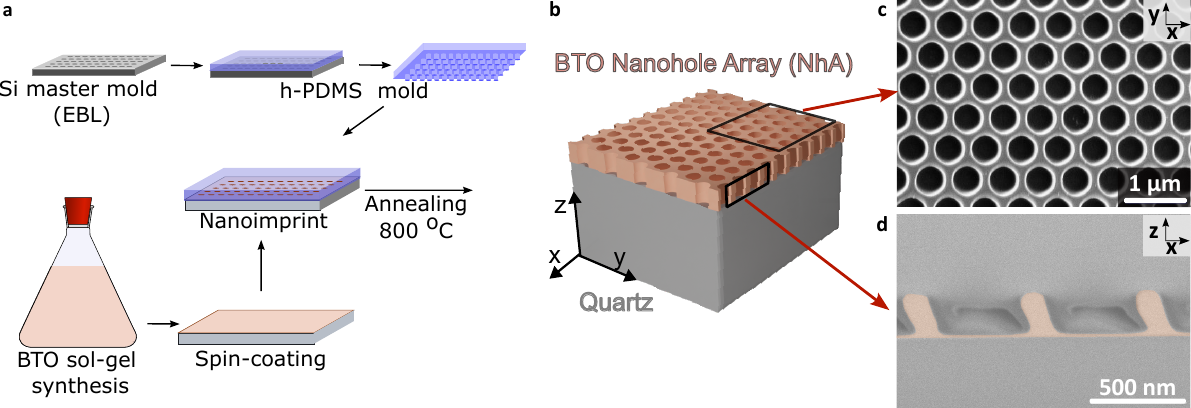}
\caption{ (a) Process flow of the direct soft nanoimprint lithorgaphy (SNIL) used for transfering the pattern from a Si master mold made with electron-beam lithography to sol-gel BTO. (b) Schematic of the resulting nanohole array in a BTO slab on a fused quartz substrate. (c) Scanning electron microscopy (SEM) image of a chromium coated BTO nanohole array in \textit{xy}-plane. (d) SEM image of the nanohole array cross-section cut in \textit{xz}-plane made with focused ion beam milling. BTO structures are shaded orange. The dark grey features between the structures are from ion beam deposition used for imaging. }
\label{fig:snilSEM}
\end{figure}

For the SNIL process depicted in \textbf{Figure} \ref{fig:snilSEM}a, a BTO sol-gel solution was prepared following a previously reported recipe for electro-optically active solution-derived BTO \cite{Edmondson2020}.

The solution was produced in ambient conditions, without a specialized glove-box and thus allows scaled-up fabrication by e.g. a roll-to-roll process \cite{Beaulieu2014}. In order to confirm the properties of the resulting material and estimate the effective refractive index for designing the nanohole arrays, a 650 nm thick characterization film was prepared alongside with the imprinted structures. X-ray diffraction measurements on the thin films confirmed the characteristic features of tetragonal and orthorhombic crystalline phases after annealing at 800 \unit{\celsius} (Figure S1a, Supporting Information). Prism coupling measurements show an effective refractive index value of 1.87 for 983 nm source (Figure S1b, Supporting Information)  which is higher than has been previously reported in nanoparticle films confirming the reduced porosity \cite{Karvounis2020}. 
Based on the measured refractive index, we design a hexagonal nanohole array with a periodicity of 612 \unit{nm}, for an estimated dielectric bandgap edge at 1500 nm,  to be imprinted into the liquid state sol-gel BTO. Transparent fused quartz was used as a substrate to enable linear and nonlinear characterization in transmission configuration. SEM images of the annealed structures retained the designed periodicity while shrinkage was observed (Figure \ref{fig:snilSEM}c \textit{xy}-plane and \ref{fig:snilSEM}d \textit{xz}-plane) as has been reported previously for direct imprinting into sol-gel \cite{Modaresialam2021}. We observe smooth side-wall surface and a uniform material filling in the mold from the imaged cross-section in Figure \ref{fig:snilSEM}d. By optimizing the spin-coating parameters, the residual film could be reduced to only 20 \unit{nm}.  The approximately 15\unit{\degree} tilt in the structures result from the slight shift when placing the mold manually on the spin-coated sol-gel solution. We achieved an aspect ratio close to 5 (570 \unit{nm}/120 \unit{nm}) with critical dimensions as small as 120 nm (SI. Figure S2). The nanohole arrays used for further optical characterization have a period of 612 nm, radius of 210 nm and thickness of 570 nm.

\subsection{Reflection and Transmission in the Photonic Crystal and Metasurface Configuration}
\begin{figure}[ht!]
%\centering
\centering\includegraphics[width=1\textwidth]{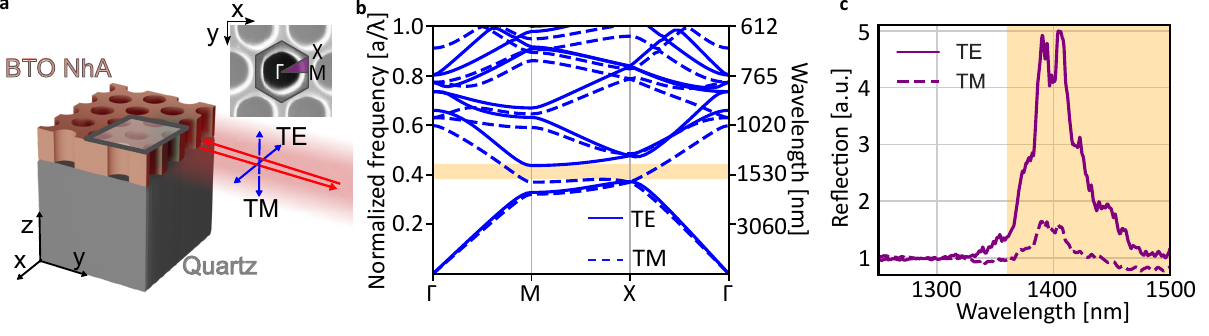}
\caption{ (a) Reflection measurement configuration with the first Brillouin zone shaded grey in the inset. Electric field polarization vector is shown for the TE and TM coupling modes. (b) Calculated energy band diagram for the structures in Figure \ref{fig:snilSEM} with the TE photonic bandgap highlighted yellow. (c) Measured reflection spectra of the slab PhC in \unit{M-X} light propagation direction with the highlighted yellow region corresponding to (b). Both TE and TM reflection spectra are normalized to measurements from an unstructured area.}
\label{fig:bandgap}
\end{figure}

The evaluation of the slab reflection and transmission spectra of the nanohole array enables to illustrate the interesting duality of PhC and metasurface behaviour, where a periodic structure supports several discrete resonance modes\cite{Qiao2018}. We use a supercontinuum white light laser and a spectrometer to experimentally show these eigenmodes at different illumination configurations (details included in the Experimental Section and Supporting Information).\\
First, we simulate and observe a PhC optical bandgap by measuring the reflection spectrum for in-plane light propagation (illumination configuration in \textbf{Figure} \ref{fig:bandgap}a, setup schematic in Figure S3, Supporting Information) from a sample cleaved along the \textit{x}-axis (along \unit{\Gamma}-M reciprocal axis). We implemented the 2D plane wave expansion method \cite{Minkov2020} with the geometrical parameters obtained from SEM images to estimate the energy band structure (Figure \ref{fig:bandgap}b). Our measurements show a strong reflection band with transverse electric (TE) polarization in the calculated wavelength range corresponding to the bandgap (Figure \ref{fig:bandgap}b, \ref{fig:bandgap}c), bandgap highlighted in orange).  No bandgap was expected for the transverse magnetic (TM) polarization guided in the film and, indeed, a significantly suppressed reflection (Figure \ref{fig:bandgap}b, \ref{fig:bandgap}c) was detected. 
 
The oscillation in the reflection can be attributed to the Fabry-Pérot resonances formed between the different imprinted nanohole array sets in the BTO slab which have a distance of 25 \unit{\micro m}(Figure S3, Supporting Information). These reflection measurements coincide well with the simulations and show a strong TE bandgap in the IR range. The ability to fabricate PhC with optimized bandgap range is important for further research into more complex devices such as BTO PhC waveguides and high Purcell factor dielectric resonators in the VIS range.  

\begin{figure}[ht!]
%\centering
\centering\includegraphics[width=1\textwidth]{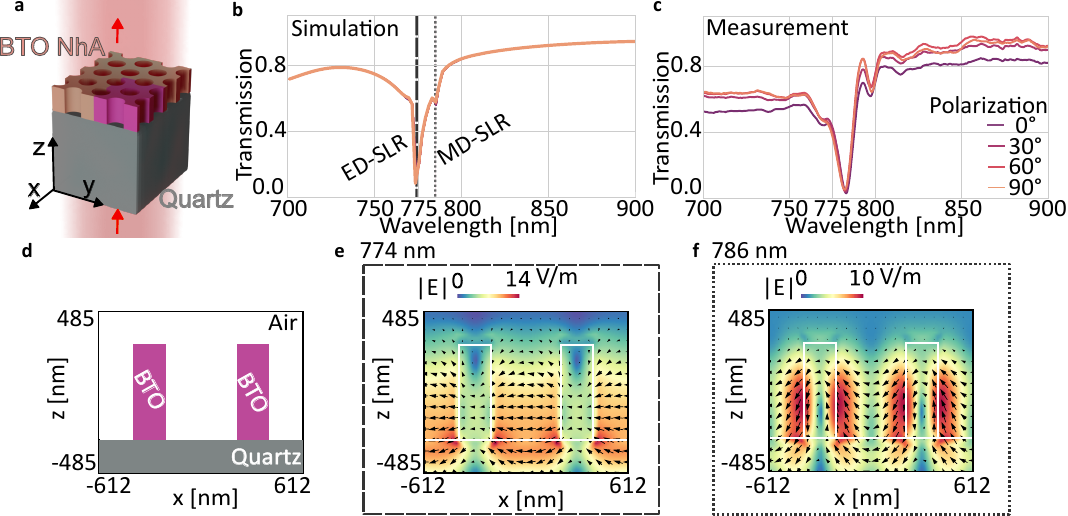}
\caption{ (a) Schematic of the hexagonal nanohole array with excitation along the z-axis direction. The red shading represents the probing laser. The unit cell used for further analysis of the fields is indicated with the purple highlighted area. Polarization independent transmission analysis: (b) simulated transmission spectra with an overlay of 0\unit{\degree} to 90\unit{\degree} polarization, (c) measured polarization independent transmission of the nanohole array. (d) Schematic of the material distribution in the \textit{xz}-plane cross-section used in part (e) and (f).  (e) Simulated absolute electric field at 774 \unit{nm} in the \textit{xz}-plane. (f) Simulated absolute electric field at 786 \unit{nm} in the \textit{xz}-plane. Arrows represent the real part of the electric field in the \textit{xz}-plane. The cross-sections are taken from the middle of the unit cell in the \textit{xz}-plane. The plane wave source polarization is kept parallel to the \textit{x}-axis. }
\label{fig:linear}
\end{figure}

To characterize the nanohole array metasurface properties, an equivalent structure was probed perpendicularly to the 2D array to excite coupled Mie resonances within the air holes or in the BTO slab (\textbf{Figure} \ref{fig:linear}a).

Finite-difference time-domain (FDTD) simulations show two distinct resonances (Figure \ref{fig:linear} b): at 774 \unit{nm} and at 786 \unit{nm}. The model was built based on parameters inferred from the SEM images and estimate of the refractive index: a period of 612 \unit{nm}, a radius of 210 \unit{nm}, a height of 570 \unit{nm}, a constant refractive index of 1.97 between 700 and 900 \unit{nm}. As the cross-section images show (Figure \ref{fig:snilSEM}d), the packing density in the structures is high thus an increased effective refractive index compared to the thin films is expected. The transmission spectra match well with the simulation results and only a slight red shift was observed experimentally with detected resonances at 782 nm and 797 nm  (Figure \ref{fig:linear}c). The strongest extinction measured at 782 nm has a quality factor of 78. This is lower than the state-of-the-art nonlinear metal-oxide resonators with 3 orders of magnitude higher quality factor values shown in lithium niobate \cite{Kang21} but further design optimization and methods to achieve free-standing or embedded structures for symmetric field distribution are expected to considerably improve these parameters. We also confirm the poly-crystalline nature of the nanostructured BTO by a polarization independent extinction at resonance in simulations as well as in the measured spectra (Figure \ref{fig:linear}b and \ref{fig:linear}c, respectively).

Analysis of the simulated electric and magnetic field profiles (Figure S4, Supporting Information) within the metasurface at the two resonances enables us to determine the different excited modes (the \textit{xz}-plane cross-section schematic annotated in Figure \ref{fig:linear}d). At higher energy resonances (Figure \ref{fig:linear}e), 774 \unit{nm} in FDTD simulation and 782 \unit{nm} in measurements, the electric field is strongly confined into the air holes with field lines along the polarization direction resembling a coupled electric dipole in a surface lattice resonant mode (ED-SLR) \cite{Conteduca2021, Castellanos2019}.  At higher wavelength resonances Figure \ref{fig:linear}f, 786 \unit{nm} in FDTD simulation and 797 \unit{nm} in measurements, strong circulating electric fields with minima in the middle are characteristic to collective coupling of a magnetic dipole and thus can be attributed to magnetic dipole surface lattice resonant mode (MD-SLR) (Figure S4f, Supporting Information) \cite{Castellanos2019}. The strongest electric field enhancement in the nanohole array is observed at the ED-SLR with up to 14 \unit{Vm\textsuperscript{-1}} field in air holes and a mean average field enhancement by 5 in the BTO strucures compared to same thickness thin film (Figure \ref{fig:linear}e). More advanced designs are expected to increase or optimize the local field confinement even further and can be adapted depending on the desired application \cite{Hentschel2023, Conteduca2021, Minkov2020}.

%\newpage
\subsection{Nonlinear signal enhancement in metasurface}

\begin{figure}[ht!]
%\centering
\centering\includegraphics[width=1\textwidth]{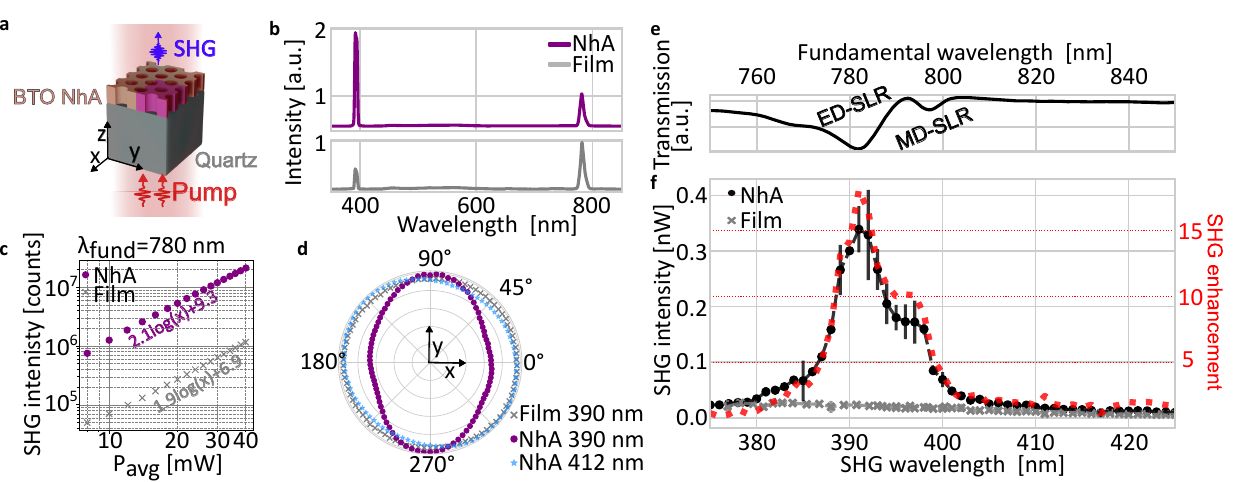}
\caption{ (a) Schematic of the SHG characterization setup. (b) Measured SHG spectra with fundamental laser at 780 nm reduced by a band pass filter (10\textsuperscript{-5} reduction at pump). (c) SHG intensity dependence on the excitation power for the nanohole array and film at 780 nm excitation. (d) Polarization dependent SHG intensity for the sol-gel thin film at 390 nm, the nanohole array at 390 nm and off-resonance (412 nm) emission. All measurements are normalized to the maximum of each measurement set. (e) Linear transmission spectrum from Figure \ref{fig:linear}(c) rescaled for comparison. (f) Pump wavelength dependent SHG emission for the film (grey markers) and the nanohole array (black markers) at an average laser power of the fundamental pump at 50 mW. The SHG enhancement by the nanostructures (P\textsubscript{NhA}/P\textsubscript{Film}) is plotted in the same graph in red.  Deviations are determined from 3 separately aligned measurements.  }
\label{fig:SHG}
\end{figure}

Taking advantage of the large transparency window and the nanohole array enabled field enhancement, we investigate here the SHG of solution derived BTO in the near-UV (375-425 \unit{nm}).
Additionally to XRD measurements (SI Figure S1), the formation of the tetragonal crystal phase was further confirmed by the spectral characterization of SHG. The BTO film and nanohole array were probed in the transmission configuration as illustrated in \textbf{Figure} \ref{fig:SHG} a with a Ti-sapphire laser (140 \unit{fs}, FWHM 10 \unit{nm}, setup details in the Experimental Section). The spectrum of the SHG in the sol-gel thin film and the PhC at 780 nm excitation (at the ED-SLR as in Figure \ref{fig:linear}b) is shown in Figure \ref{fig:SHG}b. A further evaluation of the SHG in sol-gel BTO films and imprinted structures confirmed the expected quadratic power dependence in the sol-gel derived BTO film and the nanohole array (Figure \ref{fig:SHG}c). 
We observed that a poly-crystalline material without a preferred SHG orientation is achieved for all fundamental laser excitation polarizations along the thin film plane (Figure \ref{fig:SHG}d, grey marker). 
In this poly-crystalline sample, the overlap of dipolar and quadrupolar polarization dependent SHG intensity plots is effectively averaged to an isotropic SHG when different orientations of the crystals are probed simultaneously. 
A similar homogeneous polarization independent SHG intensity was observed in the PhC at an arbitrary wavelength (Figure \ref{fig:SHG}d,  blue marker) while polarization dependent SHG was observed at the resonance (Figure \ref{fig:SHG}d,  purple marker). Simulations show that a polarization along the \textit{y}-axis forms a ED-SLR with a weaker dipole also inside the BTO slab resulting in increased maximum electric field (Figure S4c, Supporting Information), which explains the polarization dependence of the SHG at this resonance. A wavelength dependent SHG intensity was observed only in the nanohole array structures with an 18 times increased conversion efficiency at the ED-SLR and a 10 times increase at the MD-SLR (Figure \ref{fig:SHG}e and \ref{fig:SHG}f). 
The SHG shows a similar enhancement magnitude as is expected from the electric field confinement in Figure \ref{fig:linear}e and \ref{fig:SHG}f.
Our results show that the proposed method of combining the sol-gel derived BTO and nanostructures allows to considerably increase the SHG efficiency.

% \subsubsection{}

% \threesubsection{First lowest-level subsection}

\section{Conclusion}
In summary, we introduce a highly scalable method of combining sol-gel chemistry and soft nanoimprint lithography to produce nonlinear metal-oxide nanohole arrays. In addition to previously reported electro-optic properties \cite{Edmondson2020}, we show that the sol-gel derived BTO exhibits second-order optical nonlinearity. The sol-gel forms an isotropic poly-crystalline material without preferred grain orientation which averages the overall effective \unit{\chi}$^{(2)}$ tensor. Nevertheless, the random grain orientation has been shown to be beneficial for random quasi phase-matching where the SHG scales linearly with the propagation volume \cite{Muller2021}. The resulting independence from coherence length makes sol-gel BTO, and other solution processed non-centrosymmetric metal-oxides, an easily scalable material platform for broadband frequency conversion optical components. 
%Similarly to nanoparticles, sol-gel xerogels can be prepared very cost-effectively thus increasing the propagation volume to overcome the decreased effective nonlinearity is possible.
\\
Additionally, we show that solution-based processing enables to produce high aspect ratio and smooth side-wall nanostructures that exhibit PhC as well as metasurface characteristics, depending on light propagation direction. This duality in optical properties enables to design multifunctional optical devices that can control routing in-plane via e.g. photonic crystal waveguides \cite{Li2019} and meanwhile also locally couple in external radiation by behaving as nanoantennas \cite{Hentschel2023}.  We have experimentally shown the theoretically predicted enhancement in electric field confinement by over 18 times increased SHG efficiency at the resonance. Combining out-of-plane illumination induced coupled resonances with controlled slow light propagation in nonlinear material slab can potentially improve the frequency conversion efficiency further. Moreover, the large transparency window of BTO and the freedom to structure it on transparent insulating substrates enables to utilize the NUV and VIS spectrum range with minimized damage from absorption at high optical powers. The increased field in the electro-optic medium or in the nanohole cavities can be particularly beneficial for high-speed beam steering metasurfaces \cite{Sokhoyan2023} or advanced nanohole array sensors \cite{Conteduca2021}. This high-throughput fabrication platform is advantageous for improving the fabrication flexibility required for application optimized nonlinear photonic devices.

\section{Experimental Section}
\threesubsection{\quad Barium Titanate Sol-gel Synthesis}\\
Solution processed BTO sol-gel precursor was made following the procedure described by \textit{Edmunson et. al.} \cite{Edmondson2020}. Details of the exact chemicals applied in the synthesis are listed in the Supporting Information. 
0.511 \unit{g} of barium acetate is diluted in 10 ml of glacial acetic acid by stirring for an hour. 5 \unit{\celsius} water bath is prepared for combining 426 \unit{\micro l} of acetylacetate with 592  \unit{\micro l} titanium isopropoxide. Barium acetate solution in acetic acid is added to make a 1:1 stoichiometric ratio of Ba:Ti in a 0.2 \unit{\textsc{m}} solution of the metal cations. The solution is stirred for 24 h before 1:1 volume ratio dilution in anhydrous methanol to make a 0.1 \unit{\textsc{m}} solution that is used for imprinting.

\threesubsection{Hybrid-PDMS Mold Preparation and Direct Imprinting}\\
The master mold for SNIL was prepared using electron beam lithography and dry reactive ion etching process. For the hybrid PDMS mold, hard-PDMS (hPDMS, Gelest Inc.) was first spin-coated on the Si master mold followed by embedding the whole mold in soft-PDMS (Sylgard\textsuperscript{TM} 184 Elastomer Kit) to form a 2-4 mm thick mold. The hybrid PDMS was cured at 65 \unit{\celsius} for 72 hours before separating from the master mold. Mold preparation details are included in Supporting Information.
20 \unit{\micro l} of BTO sol-gel dilution was spin-coated (950 rpm, 1000 rpm/s, 5 sec) on plasma treated fused quartz substrate with the hybrid PDMS mold placed on top directly after. The mold with a 5 g weight placed on top was left to cure at 60 \unit{\celsius} for 15 hours before removal. The chip was annealed in ambient atmosphere to 800 \unit{\celsius} with heating and cooling rate of 2 \unit{\celsius min\textsuperscript{-1}}.

\threesubsection{Slab Reflection Measurement}\\
Slab reflection measurements were done in a custom microscopy setup for probing a cleaved sample with structured areas close to the cleaved edge. Broadband IR light was focused and collected from the sample edge with 20x objective (NA=0.22, Zeiss)  and the output light was coupled to a spectrometer. Focal point of the excited area was confirmed with an additional camera visualizing the sample top-down perpendicularly to the surface and beam path. Details of the setup and can be found in Figure S3, Supporting Information.

\threesubsection{Transmission Characterization}\\
Transmission measurements were done in a collinear microscopy setup with light focused on the nanohole array with a lens (LA1131 f=50\unit{mm} , Thorlabs), collected from the crystal area by a 100x objective (NA=0.75, Zeiss) and the output light was coupled to a spectrometer (Andor Schamrock, Andor Newton CCD) via a reflective collimator and a multimode fiber. 

\threesubsection{Simulations}\\
The nanohole array design was optimized using FDTD simulations (Ansys Lumerical) and the design was iterated to account for fabrication changes due to structure transfer using the PDMS mold and sol-gel shrinkage during solvent evaporation. Periodic Bloch boundaries were used in the \textit{x}- and \textit{y}-plane while perfectly matched layer boundary condition was used for both extremes of the \textit{z}-plane. Field monitors were placed in the middle of the slab in \textit{xy}-plane and through the middle of the nanoholes in \textit{xz}-plane. Electric field and magnetic field absolute and vector plots at the resonances are included in the Figure S4, Supporting Information.

\threesubsection{Second Harmonic Generation Characterization}\\
SHG was characterized in transmission using a custom built microscopy set-up with a Ti:Sapphire laser source where the laser was focused on the sample with a lens (LA1131 f=50 \unit{mm}, Thorlabs) and signal was collected with a 100x objective (NA=0.8, Olympus). The signal was captured in a sCMOS camera (Andor Zyla sCMOS) equipped with two short-pass filters (FBG39, Thorlabs) for measuring the second harmonic output (one filter for spectrum characterization). Setup and signal processing details are described in published work \cite{Timpu2019, Saerens19} and included in the Supporting Information. 

\end{onehalfspace}
\medskip
\textbf{Supporting Information}
Supporting Information is available from the author.
\textbf{Funding sources}
This work was supported by the Swiss National Science Foundation Grant 179099 and  150609, the European Union’s Horizon 2020 research and innovation program from the European Research Council under the Grant Agreement No. 714837 (Chi2-nano-oxides) and No. 862346 (PolarNon). H.W. and U.T. acknowledge financial support from the Physics Department at ETH Zurich.

% Acknowledgements
\medskip
\textbf{Acknowledgements} \par %delete if not applicable))
The authors acknowledge ScopeM and D-MATL X-Ray Service Platform at ETH Zurich for providing instrumentation for characterization, nanofabrication support from the operation team of the Binning and Rohrer Nanotechnology Center (BRNC) and the operation team of FIRST—Center for Micro and Nanoscience at ETHZ. 

\medskip

%\bibliographystyle{MSP}
%\bibliographystyle{numeric}
%\bibliography{BTOPhCpaper.bib}
%\printbibliography
\bibliographystyle{unsrt}%Used BibTeX style is unsrt
\bibliography{BTOPhCpaper}
%\bibliography{references}

%\section*{Table of Contents}
% A direct nanoimprint lithography is proposed as a scalable fabrication platform of nonlinear barium titanate photonic devices. A 2D nanohole array is produced and characterized to confirm behaviour as a 2D photonic crystal and a nonlocal resonant metasurface with over order of magnitude second harmonic generation enhancement.
%\begin{figure}[ht!]
%\centering
%\centering\includegraphics[width=0.3\textwidth]{FiguresF/coverimage.png}

%\label{fig:cover}
%\end{figure}

\end{document}